\documentclass[journal=jctcce, manuscript=article]{achemso}

\usepackage{chemformula} 
\usepackage[T1]{fontenc} 
 \usepackage[colorlinks=true,linkcolor=blue,urlcolor=blue,citecolor=blue]{hyperref}
\usepackage{array}
\usepackage{lipsum}
\usepackage{bm}
\usepackage{subfigure}
\usepackage{amssymb}
\usepackage{multirow}
\usepackage{tabularx}
\usepackage{amsmath}
\usepackage{braket}
\usepackage{caption}
\usepackage{float}
	
\usepackage{ulem}
\renewcommand{\vec}[1]{\mathbf{#1}}

\SectionNumbersOn

\author{Zhandos A. Moldabekov}
\email{z.moldabekov@hzdr.de}
\affiliation{Center for Advanced Systems Understanding (CASUS), Helmholtz-Zentrum Dresden-Rossendorf (HZDR), D-02826 G\"orlitz, Germany}

\author{Mani Lokamani}
\affiliation{Information Services and Computing, Helmholtz-Zentrum Dresden-Rossendorf (HZDR), D-01328 Dresden, Germany}

\author{Jan Vorberger}
\affiliation{Insitute of Radiation Physics, Helmholtz-Zentrum Dresden-Rossendorf (HZDR), D-01328 Dresden, Germany}

\author{Attila Cangi}
\affiliation{Center for Advanced Systems Understanding (CASUS), Helmholtz-Zentrum Dresden-Rossendorf (HZDR), D-02826 G\"orlitz, Germany}

\author{Tobias Dornheim}
\affiliation{Center for Advanced Systems Understanding (CASUS), Helmholtz-Zentrum Dresden-Rossendorf (HZDR), D-02826 G\"orlitz, Germany}

\title{Non-empirical mixing coefficient for hybrid XC functionals from analysis of the XC kernel}

\abbreviations{IR,NMR,UV}

\begin{document}







\begin{abstract}

We present an analysis of the static exchange-correlation (XC) kernel computed from hybrid functionals with a single mixing coefficient such as PBE0 and PBE0-1/3. 
We break down the hybrid XC kernels into the exchange and correlation parts using the Hartree-Fock  functional, the exchange-only PBE, and the correlation-only PBE. This decomposition is combined with  exact data for the static XC kernel of the uniform electron gas and an Airy gas model within a subsystem functional approach. This gives us a tool for the nonempirical choice of the mixing coefficient at ambient and extreme conditions. Our analysis provides physical insights into the effect of the variation of the mixing coefficient in hybrid functionals, which is of immense practical value. The presented approach is general and can be used for other type of functionals like screened hybrids. 
\end{abstract}

\begin{figure}[H]
    \centering
    \includegraphics[width=0.4\textwidth]{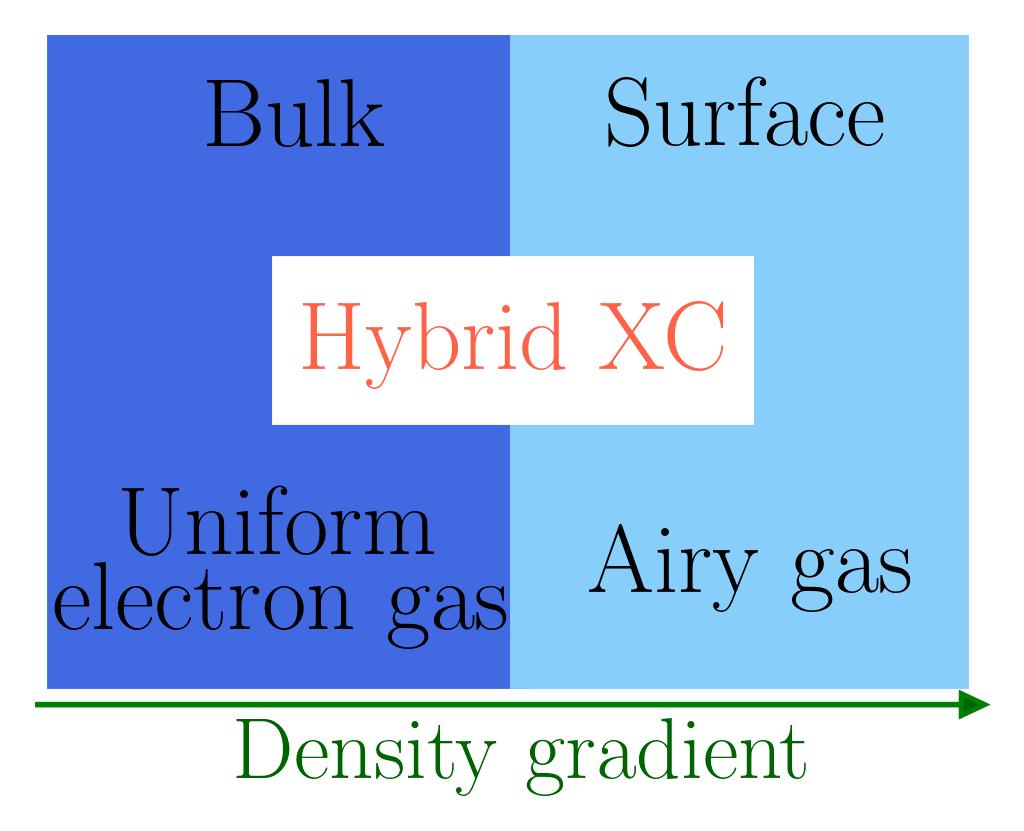}
    \caption*{\textbf{Table of Contents (TOC)/Abstract graphic.}}
    \label{fig:my_label}
\end{figure}
Modern density functional theory (DFT) based on the Kohn-Sham scheme~\cite{Jones_RMP_1989,Jones_RMP_2015} is the most widely used electronic structure method and is routinely applied in computational chemistry, condensed-matter physics, materials science, and related disciplines. The accuracy of a DFT calculation depends strongly on the particular choice of the exchange--correlation (XC) functional~\cite{Burke_JCP_2012}. It needs to be approximated and supplied as an external input to any DFT calculations.
In this regard, the generalized Kohn-Sham~\cite{KS65} density functional (KS-DFT) theory employing orbital-dependent XC functionals has a distinguished practical value \cite{RevModPhys.80.3, PhysRevX.10.021040}. Furthermore, it is one of the most promising directions in quantum chemistry aiming to develop DFT towards predictable capability.
Commonly used orbital-dependent XC functionals belong to the class of hybrid functionals which mix the exact Hartree-Fock (HF) exchange with standard density functionals, e.g,  on the level of the local-density approximation (LDA) and the generalized-density approximation (GGA) \cite{Becke, Adamo1999}.
The success of hybrid XC functionals stems from the reduction of the infamous self-interaction (or delocalization) error, which often hinders explicit density XC functionals like LDA and GGA from reaching chemical accuracy~\cite{Cohen_Science}.
Additionally, the hybrid XC functionals allow one to treat  thermal XC effects~\cite{karasiev_importance,kushal} more accurately than usual ground-state functionals due to the inherent use of thermally smeared occupation numbers~\cite{hybrid_results, PhysRevLett.118.225001, PhysRevLett.126.025003}. This is particularly important for the so-called warm dense matter (WDM) \cite{dornheim_physrep18_0, wdm_book,new_POP}, which is a state of matter generated for example by laser heating or shock compression at facilities such as the European X-ray Free-Electron Laser (XFEL) \cite{Tschentscher_2017} and Linac coherent light source (LCLS) at SLAC \cite{doi:10.1126/sciadv.abo0617}, and naturally occurs in astrophysical objects like the interior of giant planets \cite{Benuzzi_Mounaix_2014, doi:10.1126/sciadv.abo0617}.
Important technological applications of WDM include the discovery of novel materials at extreme conditions~\cite{Lazicki2021, Kritcher2020, Kraus2017} and hot-electron chemistry \cite{Brongersma2015}.
Therefore, the rigorous understanding of hybrid XC functionals within the generalized KS-DFT approach is of immense importance for physics and quantum chemistry at both ambient and extreme conditions. 

Grounded upon a rigorous \textit{ab initio} relation known as the adiabatic connection formula for the exchange-correlation energy \cite{Becke, PhysRevX.10.021040}, hybrid XC functionals are typically  constructed by admitting a certain amount of the HF exchange into the total exchange part of the XC functional.  
Following Perdew, Ernzerhof, and Burke \cite{doi:10.1063/1.472933}, often hybrid XC functionals are represented as
\begin{equation}\label{eq:pbe0a}
    E_{\rm xc}[\rho_{\sigma},n]=E_{\rm c}^{\rm DF}[n]+aE_{x}^{\rm HF}[\rho_{\sigma}]+(1-a)E_{\rm x}^{\rm DF}[n]\ ,
\end{equation}
where $a$ is referred to as the mixing parameter and DF stands for density functional.  Examples for Eq.~(\ref{eq:pbe0a}) are given by PBE0~\cite{Adamo1999} with $a=1/4$  and PBE0-1/3~\cite{Cortona2012} with $a=1/3$ , where  $E_{\rm xc}^{\rm DF}=E_{\rm xc}^{\rm PBE}$~\cite{PBE}. To ease the computational cost of PBE0, a screened version of its exchange part was implemented in HSE03  \cite{Heyd2003} and HSE06 \cite{Heyd2006, Krukau2006}, where the Coulomb interaction $\sim 1/r$ is substituted by a screened interaction $\sim {\rm erfc}(\Omega r)/r$, with $\Omega$ being the screening parameter. Arguably, for solids, these four are the most commonly applied and most famous hybrid XC functionals. Further, we refer to the hybrid functionals with an arbitrary mixing coefficient $a$ in Eq.~(\ref{eq:pbe0a}) and with  $E_{\rm xc}^{\rm DF}=E_{\rm xc}^{\rm PBE}$ as the PBE0-type functionals.

While a general framework of the KS-DFT with hybrid XC functionals has been rigorously derived by Garrick \textit{et al.}~\cite{PhysRevX.10.021040}, the choice of the mixing degree for solids had remained somewhat intuitive and justified empirically by computing material properties and comparing them with experimental measurements such as lattice constants,
bulk moduli, the vacancy formation energy, and to atomic data.
The usually quoted rational for the choice of the mixing coefficient is due to Perdew \textit{et al.} \cite{doi:10.1063/1.472933}, who used an analysis of atomization errors of typical molecules from the M{\o}ller-Plesset perturbation expansion to conjecture $a=1/4$.
 Clearly, there is ample room left for the variation of the mixing degree and of the screening parameter for systems different from molecules.  The situation is particularly unsatisfactory for WDM, where such basic atomic properties like an atomization energy become ill defined due to the smearing of the boundary between bound and free states at high  pressure or temperature\cite{Bohme_PRL_2022, Hu2022, PhysRevResearch.3.023026, Hu2020}, and where bulk properties such as the vacancy formation energy or bulk moduli cannot be accurately measured due to the extreme conditions. Moreover, bulk properties of materials like the lattice constant (interparticle distance/separation), stress tensor etc., at high pressures and temperatures  often differ significantly from those at ambient conditions. This calls for an approach that allows to rationalize the choice of the mixing coefficient and screening without employing properties of individual atoms and molecules or other physical properties limited to ambient conditions. At the same time, it is preferable that such a rational has some connection to properties that are well defined and measurable in experiments across temperature and pressure regimes. 
  Here we show that the static XC kernel $K_{\rm xc}(q)$ [were $q$ is a wavenumber, see Eq.~(\ref{eq:chi}) below] can serve this purpose. First we discuss  $K_{\rm xc}(q)$ of the PBE0-type hybrid functionals by comparing it to archetypal electron gas models~\cite{giuliani2005quantum} to illustrate the concept. After that we describe how this can be related to the static density response  function (susceptibility) $\chi(q)$ that can be measured in experiments via the x-ray Thomson scattering (XRTS) technique~\cite{siegfried_review,Dornheim2022Physical}.
  
  To begin with, we argue that the choice of the mixing coefficient can be based on $K_{\rm xc}(q)$ of the uniform electron gas (UEG)~\cite{giuliani2005quantum,loos} valid for bulk systems  and on $K_{\rm xc}(q)$ of the AM05 functional~\cite{PhysRevB.72.085108} that incorporates the limit of the Airy gas at large gradients like in near-surface regions.
  This is demonstrated for the static XC kernel of the PBE0-type functionals by varying the mixing coefficient in a wide range of values. 
  The AM05 incorporates the exact exchange of the  Airy gas and has the correlation part that is fitted to reproduce jellium surface energies (since exact quantum Monte Carlo data for the Airy gas are not available).
  Importantly, the AM05 enables an accurate and universal treatment of systems with electronic surfaces \cite{Mattsson, PhysRevB.72.085108}. Our choice of UEG and jellium surface based reference data for $K_{\rm xc}(q)$ is motivated by the fact that the functionals based on these generic-model systems are more widely applicable for solids than the XC functionals fitted to specific materials (see, e.g., Refs. \cite{Paier, PBE, Mattsson, PhysRevB.72.085108, Vitos1}).

First, we recall that the complete information about the density response of a given system of interest is contained in the dynamic density response function, which can be conveniently expressed for homogeneous systems as~\cite{giuliani2005quantum},
\begin{eqnarray}\label{eq:kernel}\label{eq:chi}
 \chi(\mathbf{q},\omega) = \frac{\chi_0(\mathbf{q},\omega)}{1 - \left[v(q)
 +K_\textnormal{xc}(\mathbf{q},\omega)\right]\chi_0(\mathbf{q},\omega)}.
\end{eqnarray}
Here $v(q)=4\pi/q^2$ is the Coulomb potential in reciprocal space and $\chi_0(\vec q, \omega)$ denotes a known reference function such as the response of the ideal Fermi gas in the case of a UEG, or the KS response function $\chi_\textnormal{KS}(\mathbf{q},\omega)$ for linear-response time-dependent DFT (LR-TDDFT) calculations for real materials~\cite{marques2012fundamentals}.
The exact density response is then given by the combination of $\chi_0(\mathbf{q},\omega)$ with the dynamic XC kernel $K_\textnormal{xc}(\mathbf{q},\omega)$.

The XC kernel can be computed by inverting Eq. (\ref{eq:chi}) if the density response function is known. To compute the static XC kernel $K_\textnormal{xc}(\mathbf{q})=K_\textnormal{xc}(\mathbf{q},\omega=0)$, we perturb the UEG with an external static harmonic perturbation $V_{\rm ext}=2A\cos (\vec q \cdot \vec r)$. Then, $\chi(q)=\chi_0(\vec q, \omega=0)$ is found from the difference $\delta n(\vec r)$ between the perturbed and unperturbed densities \cite{PhysRevLett.75.689, dft_kernel,Dornheim_PRR_2021}:
  \begin{equation}\label{eq:dn}
    \delta n(\vec r)=2A \cos(\vec q \cdot \vec r) \chi(\vec q).
\end{equation}

This approach to compute $\chi(q)$  was used for the UEG by Dornheim {\textit{et al}} \cite{PhysRevLett.125.085001, PhysRevE.96.023203} within path integral Monte Carlo (PIMC) method. Recently, this method has been extended to the KS-DFT and has been illustrated for the UEG and warm dense hydrogen \cite{dft_kernel}. Importantly, we circumvent the need to compute the second order functional derivatives of the XC functionals and, thus, we can calculate the static XC kernel across Jacob's Ladder \cite{dft_kernel}. The key ingredient of our analysis is the exact quantum Monte Carlo data for the so-called local field correction (LFC) of the UEG, which is connected to the XC kernel as $G(q)=-K_{\rm xc}(q)/v(q)$ and is commonly used in the quantum theory of the electron liquid \cite{giuliani2005quantum,kugler1,Dornheim_PRL_2020_ESA}.

The UEG is conveniently described by the mean-inter particle distance $r_s$ and the reduced temperature $\theta=T/T_F$, where $r_s$ is in units of the Bohr radius and $T_F$ is the Fermi temperature (energy) \cite{doe-report-17,Ott2018}. All results are presented in Hartree atomic units. We consider metallic densities and set $r_s=2$, which is typical for both solids and WDM \cite{PhysRevB.103.125118}. The results are presented for the ground state $\theta=0.01$ with strong electron degeneracy and for the case with partial electron  degeneracy $\theta=1$, i.e. $T=T_F$. The former case corresponds to ambient conditions and the latter case is characteristic for the WDM state.
We note that the perturbation amplitude $A$ in Eq. (\ref{eq:dn}) must be small enough to avoid higher order non-linear effects. This aspect was extensively analyzed within the KS-DFT method for the UEG by Moldabekov \textit{et al}~\cite{Moldabekov_JCTC_2022}. Further, we drop  vector notation for the wavenumber and set $\vec q$ in Eq. (\ref{eq:dn}) to be along the $z$-axis.

The ABINIT package \cite{Gonze2020, Romero2020, Gonze2016, Gonze2009, Gonze2005, Gonze2002} has been used for the KS-DFT simulations of the perturbed free electron gas. 
The results are presented with $N=38$ electrons in the main-cell at $\theta=0.01$ and with $N=14$ electrons at $\theta=1.0$. We note that finite-size effects have been extensively investigated in the literature~\cite{Chiesa_PRL_2006,dornheim_prl,Dornheim_JCP_2021,dornheim_ML, Moldabekov2022, Moldabekov_JCP_2021, Moldabekov_PRB_2022} and are expected to be small at these conditions.
The corresponding cell lengths are $L=10.839~{\rm Bohr}$ and $L=7.7703~{\rm Bohr}$, respectively. The perturbation wavenumber is defined by the length of the main simulation cell as $q=2\pi j/L$, where $j\geq 1$ is a positive integer number. 
The number of bands at $\theta=0.01$ is set to $76$ and the number of bands at $\theta=1.0$ is set to $200$.
The calculations are performed with periodic boundary conditions and  a {\textit k}-point grid of $8\times 8\times 8$. Maximal kinetic energy cut-off is set to $13~{\rm Hartree}$. Self-consistent-field (SCF) cycles for the solution of the Kohn-Sham equations where converged with absolute differences in the total energy of $\delta E<10^{-7}~{\rm Ha}$. The perturbation amplitude is set to $A=0.01$.
For the detailed discussion of the convergence with respect to the simulation parameters we refer the reader to our recent paper \cite{hybrid_results, dft_kernel}.

\begin{figure*}[!t]
\includegraphics[width=0.99\textwidth]{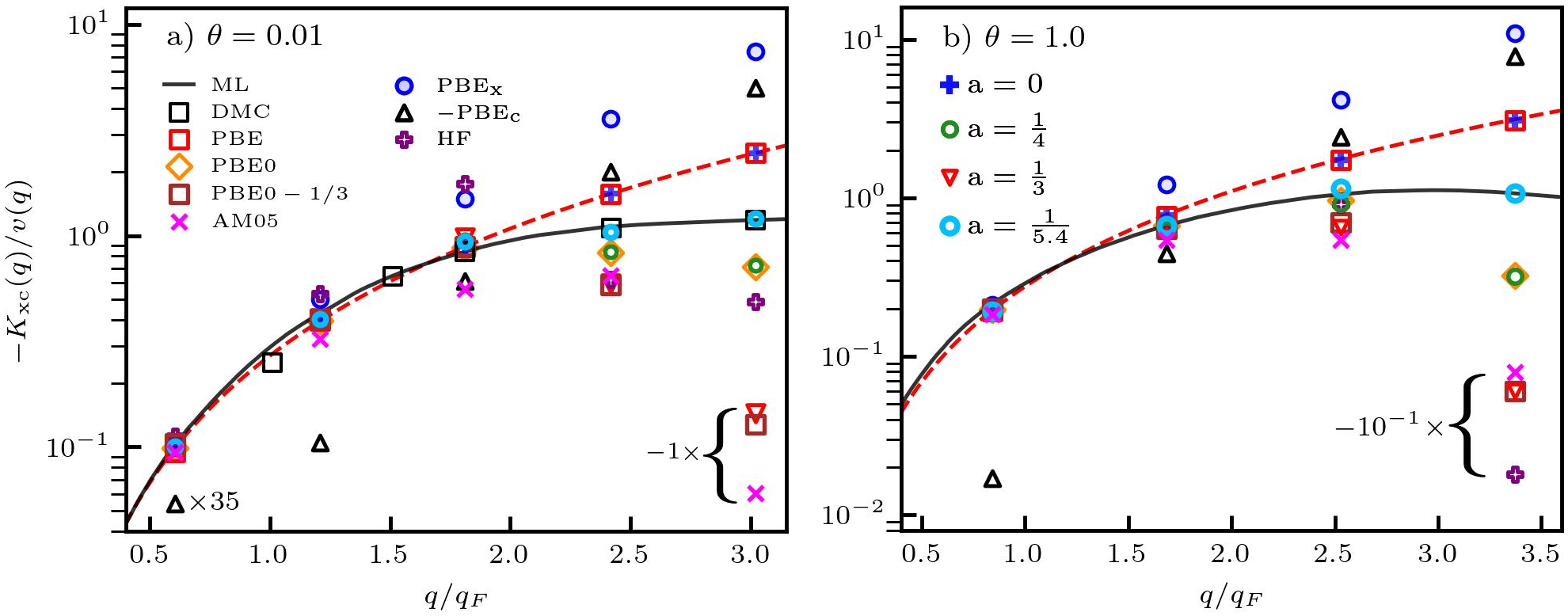}
\caption{\label{fig:kernel}  a)  Static XC-kernel of the UEG at a) $\theta=0.01$  and b) $\theta=1$ for $r_s=2$. 
The exact  diffusion quantum Monte Carlo (DMC) results by Moroni \textit{et al.} \cite{PhysRevLett.75.689} are presented by black squares for $\theta=0.01$. The solid line is the ML representation of the exact quantum Monte Carlo results by Dornheim \textit{et al.}\cite{dornheim_ML}.
The dashed line illustrates the quadratic dependence of the PBE data on $q$.
The mixing parameter $a$ indicates the data obtained by combining the results for kernels from separate HF, the exchange-only PBE (labeled as ${\rm PBE_x}$), and the correlation-only PBE (labeled as ${\rm PBE_c}$) calculations as $K_{\rm xc}(q, a)=K_{\rm xc}^{PBE_c}(q)+a K_{\rm xc}^{HF}(q)+(1-a)K_{\rm xc}^{PBEx}(q)$. Other symbols correspond to the KS-DFT data computed using PBE, PBE0, PBE0-1/3, and AM05 as it is described in the main text. 
}
\end{figure*}

The results for the $K_{\rm xc}$ of the UEG for $r_s=2$ at $\theta=0.01$ and $\theta=1.0$ are presented in Fig. \ref{fig:kernel}.
Remarkably, we find exactly the same trend for both temperatures.
In particular, we show $K_{\rm xc}$ computed  using the PBE, PBE0, PBE0-1/3, and AM05 functionals and we compare them with the exact  diffusion quantum Monte Carlo (DMC) results by Moroni \textit{et al.} (black squares) \cite{PhysRevLett.75.689} at $\theta=0.01$ as well as  with the  machine-learning representation (ML) by Dornheim {\textit {et al.}} \cite{dornheim_ML} based on an extensive PIMC simulations for the WDM parameters (solid  line). The dashed red line illustrates the quadratic $q$ dependence of the PBE based $K_{\rm xc}(q)$ according to the compressibility sum-rule  \cite{dft_kernel}. It becomes exact in the limit of $q\to0$. The results for $K_{\rm xc}(q)$ based on various LDA, GGA, meta-GGA, and hybrid XC functionals, including the data for the PBE, PBE0, PBE0-1/3, and AM05, were reported in Refs. \cite{dft_kernel, hybrid_results}. In this work, we break down the PBE0-type functionals into exchange and correlation parts to get physical insight into the effect of variation of the mixing coefficient.  For that, we performed new simulations using only the HF orbital-dependent functional, only the exchange part of the PBE (${\rm PBE_x}$), and only the correlation part of the PBE (${\rm PBE_c}$).

Let us start our discussion from the special case of Eq.~(\ref{eq:pbe0a}) with $a=0$, i.e. without the HF contribution.
In this case Eq.~(\ref{eq:pbe0a}) reduces to the PBE functional. 
From Fig. \ref{fig:kernel}, we see that the results for the kernels in ${\rm PBE_x}$  (blue circles) and ${\rm PBE_c}$ (black triangles) vastly differ by their values from the full PBE data (red squares) and all other results. Thus, we use a logarithmic scale for comparative analysis.  The $K_{\rm xc}(q)$ based on ${\rm PBE_x}$ and ${\rm PBE_c}$ have opposite signs (${\rm PBE_c}$ data is shown with a minus sign). The first numerical observation is that the ${\rm PBE_x}$ kernel is dominant over the ${\rm PBE_c}$ kernel in the entire considered $q$ range, $q\lesssim 3 q_F$ (with $q_F$ being the Fermi wavenumber).
Particularly, at $q<q_F$, the ${\rm PBE_x}$ kernel is larger than the ${\rm PBE_c}$ kernel by at least one order of magnitude in absolute value. In Fig. \ref{fig:kernel}a), the data point at $q\simeq 0.6 q_F$ for the ${\rm PBE_c}$ kernel is multiplied by a factor of 35 for a better illustration.
The direct sum of the ${\rm PBE_x}$ and ${\rm PBE_c}$ data for the UEG reproduces the results obtained using the full PBE functional since  the XC kernel is equivalent to the second order functional derivative of the XC functional and for PBE we have $E_{\rm xc}^{\rm PBE}[n]=E_{\rm x}^{\rm PBE}[n]+E_{\rm c}^{\rm PBE}[n]$ \cite{PBE}.

Next, we consider the case with $a=1/4$, which is the definition of the PBE0 (depicted using orange rhombus). 
For the PBE0, one needs a piece of the HF exchange. The full HF data (purple plus signs) are in close agreement with  ${\rm PBE_x}$ data at $q<1.5q_F$ and have a maximum at $1.5q_F<q<2q_F$. At $q>2q_F$, the HF kernel underestimates the $K_{\rm xc}(q)$ compared to the exact quantum Monte Carlo data. Now we sum the quarter of the HF kernel, the two-thirds of  the ${\rm PBE_x}$ kernel and the full ${\rm PBE_c}$ kernel according to  Eq.~(\ref{eq:pbe0a}) with $a=1/4$. The resulting data points (green circles) are in agreement with the data computed using the PBE0 functional. This confirms that the PBE0 XC kernel for the UEG can be found by adding the exchange and  correlation parts computed separately.  
From Fig. \ref{fig:kernel} we see that, on the one hand, the replacement of 1/4 of the ${\rm PBE_x}$ kernel by the quarter of the HF kernel significantly reduces the PBE0 kernel at $q>2q_F$ compared to the exact data for the UEG. On the other hand, this replacement leaves the PBE0 kernel  almost unaffected at $q<2q_F$. The reason for this is the aforementioned observation that the HF kernel is nearly the same as the  ${\rm PBE_x}$ kernel at $q<2q_F$.

Third, we consider the PBE0-1/3 based results (brown squares). The same way as we did it for the PBE0, here we use the static XC kernel data from the separate HF, ${\rm PBE_x}$, and  ${\rm PBE_c}$ based simulations to combine them according to Eq.~(\ref{eq:pbe0a}) with $a=1/3$. The result (red-inverted triangles) reproduces the data obtained from independent simulations using the PBE0-1/3 functional (see  Fig. \ref{fig:kernel}). We observe that the admission of one third of the HF exchange does not affect the PBE0-1/3 kernel at $q<2q_F$ due to the close value of the HF kernel and the  ${\rm PBE_x}$ kernel at these wavenumbers.
At the same time, the mixing of one third of the HF exchange leads to the significant reduction of the PBE0-1/3 kernel compared to the exact data for the UEG at $q>2q_F$. At $q>3q_F$, the PBE0-1/3 kernel even changes its sign with the increase in the wavenumber (indicated in Fig. \ref{fig:kernel} by multiplication by a negative factor). 
The reason for that is that the HF kernel reduces with the increase in $q$ at $q>2q_F$ and thus the difference between the exchange and correlation parts of the PBE0-1/3 kernel decreases and eventually the sign of the kernel changes at $q>3q_F$. 

Therefore, we have established that  at $q>2q_F$  the disagreement between the exact data for the UEG and both PBE0-1/3 and PBE0 increases with the increase in the wavenumber since the HF kernel decreases with $q$. 
On the other hand, we see
from  Fig. \ref{fig:kernel} that the same decreasing trend  at $q>2q_F$ is the case for the AM05 based kernel (pink crosses). In fact, the agreement of the  PBE0 and PBE0-1/3 kernels with the AM05 kernel is much better than the agreement of the PBE kernel with the AM05. This is also the case for the HSE06 and HSE03, which are screened versions of PBE0 \cite{hybrid_results}.
A relevant observation is that the AM05 performs similarly well as the PBE0 and the HSE06 and much better than PBE for solids and surfaces \cite{Mattsson}.  The success of the AM05 is due to a subsystem functional approach used to develop it \cite{sub_systemDFT}.
Within the subsystem functional approach, the AM05 mimics the Airy gas model at large scaled density gradients and the UEG model at small density gradients  \cite{PhysRevB.72.085108}. Indeed, we see that at $ q\lesssim 1.5q_F$ the AM05 based kernel is in agreement with the PBE kernel and with the exact data for the UEG. From  Fig. \ref{fig:kernel}, one can see that the increase in the mixing coefficient from $a=1/4$ (used for the PBE0) to $a=1/3$ (used for PBE0-1/3) leads to the closer agreement of the PBE0-type kernel with the AM05 kernel.

\begin{figure*}[!t]
\includegraphics[width=0.99\textwidth]{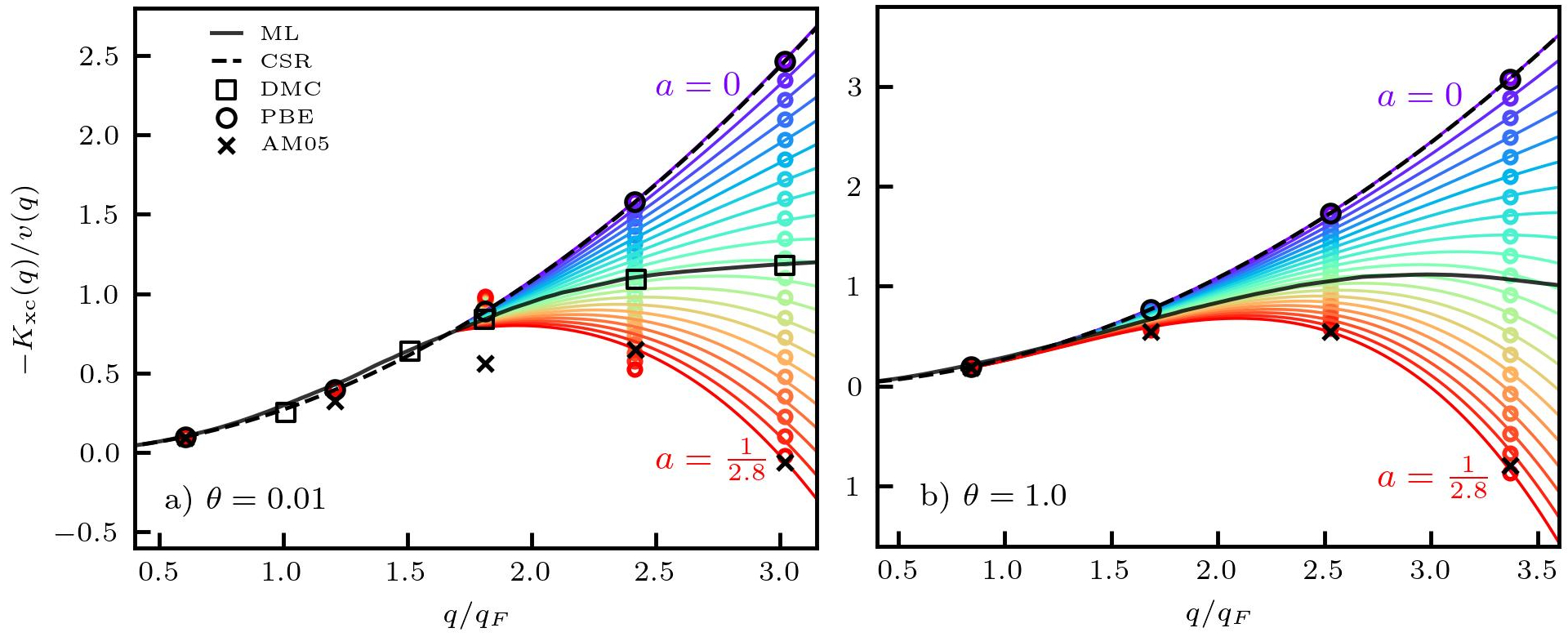}
\caption{\label{fig:a_variation}  a)  Static XC-kernel at a) $\theta=0.01$  and b) $\theta=1$ for $r_s=2$. 
The exact  diffusion quantum Monte Carlo (DMC) results by Moroni \textit{et al.} \cite{PhysRevLett.75.689} are presented by black squares for $\theta=0.01$. The solid line is the ML representation of the exact quantum Monte Carlo data by Dornheim \textit{et al.}~\cite{dornheim_ML}.
The dashed lines illustrate the quadratic dependence of the PBE data on $q$ in logarithmic scale.
The mixing parameter $a$ in Eq. (\ref{eq:pbe0a})  is varied in the range $0\leq a \leq 1/2.8$. The KS-DFT data corresponding  to different $a$ values are presented by circles with different colors and the corresponding lines depict spline interpolations between these data points.   
}
\end{figure*}

After this illustration of our approach for the analysis of the PBE0-type functionals for the examples of $a=1/4$ and $a=1/3$, we now use the HF, ${\rm PBE_x}$, and  ${\rm PBE_c}$ based kernels to  investigate the PBE0-type kernel for a wide range of the mixing parameter $a$. In particular, we achieve an excellent agreement with the exact data for the UEG at the considered wavenumbers by choosing $a=1/5.4$ as it is clearly illustrated for both $\theta=0.01$ and $\theta=1$ in Fig. \ref{fig:kernel} (light blue circles). In Fig. \ref{fig:a_variation}, we scan $a$ values from $a=0$ to $a=1/2.8$ in order to show that the corresponding PBE0-type kernel changes from the PBE kernel to the AM05 kernel with the increase in $a$ at $q>2q_F$ and remains in a good agreement with the exact data for the UEG at $q<2q_F$. 

An important conclusion from these results is that the mixing coefficient for the PBE0-type functionals can be chosen using the XC kernel from the UEG model or from the AM05 which also incorporates the Airy gas model within the subsystem functional approach at large density gradients.  These two models are archetypal for the electrons in the bulk region \cite{PBE, PBEsol} and in the surface region \cite{PhysRevB.72.085108,  Mattsson, Vitos1, PhysRevLett.81.3487} of solids. 
The functionals that are based on such generic models are referred to as \textit{nonempirical} functionals since they are valid for larger variation of the materials than functionals created for some specific materials \cite{doi:10.1063/1.472933, PBE, Mattsson}. 
In this way, we provide a tool for the construction of hybrid functionals that are nonempirical for both solid-state systems and WDM. Indeed, one could perform a similar analysis for the screened hybrid functionals by varying the screening parameter $\Omega$. Moreover, the static XC kernel of the  UEG based on the HF, ${\rm PBE_x}$, and  ${\rm PBE_c}$ functionals can be calculated over a wide range of densities and temperatures to find the optimal parameters of the hybrid XC functionals such us the mixing coefficient.

The presented approach is particularly valuable for WDM.  For example, there is no need to care about surface properties since WDM samples generated  by laser heating or shock compression do not have a clearly distinguishable surface. Thus, one could choose e.g.~$a=1/5.4$ at metallic densities to have a better description of the bulk properties.

We are convinced that our new insights into the nonempirical mixing coefficient in hybrid functionals will be useful for a number of applications both at ambient conditions and in the WDM regime. In particular, we stress that the XC kernel is very important in its own right and constitutes the key input for the computation of material properties such as effective potentials~\cite{doi:10.1063/5.0097768, PhysRevB.104.195142, zhandos2, giuliani2005quantum}, quantum fluid dynamics~\cite{Murillo,zhandos_pop18,Moldabekov_SciPost_2022, cpp_202100170}, or for LR-TDDFT simulations of real materials~\cite{marques2012fundamentals} and  for plasmonics \cite{doi:10.1063/5.0100797}.
A particularly interesting possibility is the experimental verification of our work for the hybrid XC kernel in XRTS experiments with WDM~\cite{siegfried_review}.
XRTS is a standard tool of diagnostics of WDM and measurements can be performed at both small $q<2q_F$ and large $q>2q_F$ wavenumbers \cite{Preston, GFGN2016:matter, Frydrych2020}. In fact, the XRTS signal is the dynamic structure factor of the electrons convolved with the probe function. 
Recently, it has been shown that a two-sided Laplace transform of the measured XRTS intensity can be used to compute the  imaginary-time density–density correlation function with significantly reduced experimental noise \cite{Dornheim_T_2022}. This allows one to subsequently find the static density response function $\chi(q)$ via the imaginary-time version of the fluctuation–dissipation theorem \cite{Dornheim2022Physical}. Thus, the static XC kernel that has been shown here to be a tool for the nonempirical definition of the mixing coefficient in the hybrid functionals can also be probed in XRTS measurements with matter at extreme conditions.


\section*{Acknowledgments}
This work was funded by the Center for Advanced Systems Understanding (CASUS) which is financed by Germany’s Federal Ministry of Education and Research (BMBF) and by the Saxon state government out of the State budget approved by the Saxon State Parliament. We gratefully acknowledge computation time at the Norddeutscher Verbund f\"ur Hoch- und H\"ochstleistungsrechnen (HLRN) under grant shp00026, and on the Bull Cluster at the Center for Information Services and High Performance Computing (ZIH) at Technische Universit\"at Dresden. The authors also thank Henrik Schulz and Jens Lasch for providing very helpful support on high performance computing at HZDR.

\bibliography{ref}

\end{document}